\documentclass{ws-ijbc}

\def\cropmarks{}\let\trimmarks\cropmarks

\usepackage{xcolor}
\usepackage[verbose]{hyperref}
\hypersetup{colorlinks=false,allbordercolors=blue,pdfborderstyle={/S/U/W 1}}

\begin{document}

\markboth{A. Raj and M. R. Paul}{Spatiotemporal Chaos with Extended Spatial Interactions}

\title{Spatiotemporal Chaos with Extended Spatial Interactions}

\author{A. Raj \& M. R. Paul$^*$}

\address{Mechanical Engineering Department, Virginia Tech \\ 
Blacksburg, VA 24061, USA \\
mrp@vt.edu$^*$}

\maketitle

\begin{abstract}
The chaotic dynamics of a large one-dimensional lattice of nonlinear maps with a spatial coupling of increasing extent is explored using the covariant Lyapunov vectors (CLVs). A theoretical approach is developed that predicts the spectral variation of the CLVs and the spectrum of Lyapunov exponents. The theoretical approach is independent of the details of the mapping function and requires only a description of the lattice coupling. The general structure of the spatial power spectrum of the CLVs is determined using only the eigenvectors of the coupling matrix.  This yields a decomposition of the CLVs into two regimes. The first regime, corresponding to the largest Lyapunov exponents, contains highly entangled CLVs with significant contributions from large scale spatial structures and aligns with the sorted eigenvectors, which occur in pairs with wavenumbers that increase linearly with their index. The second regime contains significant mixing of length scales with increasing index. The index separating these two regimes is approximately equal to the fractal dimension of the dynamics.
\end{abstract}


\section{Introduction}
\label{section:introduction}

Many important problems facing science and technology today are large spatially-extended systems that are driven far-from-equilibrium to exhibit complex spatiotemporal dynamics~\cite{cross:1993}. Examples include geophysical fluid dynamics~\cite{marshall:1999}, the spread of information in a social network~\cite{barthelemy:2011}, the striking patterns of reacting-advecting-diffusing systems~\cite{mukherjee:2022}, the formation of spiral galaxies~\cite{sellwood:2022}, the importance of weak mean-flows in fluid dynamics~\cite{newell:1990:jfm}, the dynamics of a pandemic in a highly connected population~\cite{song:2010}, and the interactions of billions of neurons in the human brain~\cite{yuste:2024}.

In many systems, the disorder is generated locally in space which is then distributed throughout the system by mechanisms of spatial coupling~\cite{hohenberg:1989}. There are many phenomena that can generate spatial couplings.  In thermal-fluid transport problems, spatial coupling can be the result of diffusive, reactive, convective, and magnetic phenomena to name a few. More broadly, many problems can be described as a spatial network where the types of couplings can be very rich and diverse~\cite{watts:1998,majhi:2022}.

Powerful ideas from dynamical systems theory~\cite{eckmann:1985} provide physical insights into the complex dynamics of high-dimensional chaotic systems. In particular, the covariant Lyapunov vectors (CLVs)~\cite{pikovsky:2016} provide a description of the growth or decay of small perturbations to the nonlinear dynamics. Using the dynamic algorithm of~\citet{ginelli:2007} it is now possible to compute hundreds of CLVs for large systems, including systems of nonlinear partial differential equations, with currently available computing resources.

The CLVs have been used to provide physical insights into the dynamics of a wide range of problems of varying complexity. Using CLVs, important insights into the chaotic dynamics of coupled map lattices (CMLs) have been gained~\cite{pikovsky:1998} including the role of conservation laws on chaos~\cite{puri:1991,barbish:2023}. Fundamental differences in the regimes of amplitude and phase turbulence of the complex Ginzburg-Landau have been uncovered using the CLVs~\cite{takeuchi:2011}.  CLVs have been used to investigate the chaotic dynamics of canonical model systems described by partial differential equations such as the Kuramoto-Sivashinsky equation~\cite{yang:2009}. Fluid systems have been investigated including Kolmogorov flows~\cite{inubushi:2012}, the spatiotemporal chaos of Rayleigh-B\'enard convection~\cite{xu:2016,xu:2018}, fluid turbulence~\cite{inubushi:2015}, and weather models~\cite{lucarini:2020}.

Many of the new insights provided by the CLVs can be traced to their quantification of the directions in tangent-space of the growth or decay of small perturbations to the dynamics. The CLVs have been used to determine the stable and unstable manifolds of the dynamics in state-space~\cite{ginelli:2007,takeuchi:2011,xu:2016}, to quantify the degree of hyperbolicity of the dynamics~\cite{ginelli:2007,inubushi:2012,takeuchi:2011}, to estimate the dimension of the inertial manifold~\cite{yang:2009}, to quantify the entanglement of the CLVs~\cite{barbish:2023}, and to provide a decomposition of the tangent-space into physical and transient modes~\cite{yang:2009}.

We use a large 1D lattice of coupled maps to explore fundamental questions regarding the role of spatially extended coupling on chaotic dynamics. CMLs have a rich literature and have been used to generate fundamental insights into high-dimensional chaotic dynamics. Early work explored a broad range of chaotic phenomena~\cite{kaneko:1989-stc}. For example, CMLs have been used to explore spiral patterns~\cite{sbitnev:1997-checkerboard}, nonlinear predation models~\cite{han:2024}, the dynamics of clouds~\cite{yanagita:1997}, spatiotemporal chaos and chimera states~\cite{omelchenko:2011}, and the long-distance coupling of trees through pollen exchange~\cite{satake:2002}. CMLs played a central role in the exploration of a thermodynamic description of chaos~\cite{bourzutschky:1992,miller:1993,egolf:2000:science} and the synchronization of chaos in the presence of spatial coupling~\cite{batista:2020}.  CMLs have provided insights into the nonlinear dynamics and pattern formation of inhomogeneous lattices~\cite{biswas:2016-patterns}, ecological systems~\cite{zhong:2021-spatiotemporal}, and the spread of an epidemic~\cite{salman:2024-spatiotemporal}.

The eigenvalue spectrum of the coupling matrix has long served as the primary lens for understanding the interplay between network topology and synchronization stability~\cite{gade1996synchronization,pecora1998master,gong2008stability}. This approach has proven instrumental in analyzing synchronous states in diffusively coupled lattices, revealing that stability regions for specific coupling matrices can remain invariant even as the system size increases~\cite{hu1998instability}. Such findings underscore the pivotal role of the coupling operator in different types of couplings that govern the spatiotemporal dynamics of CMLs~\cite{jost2001spectral,zhang2018spatiotemporal}.

Beyond the nature of the coupling operator, the physical scale of interaction significantly affects system behavior~\cite{stahlke2011length}. In this context, extended spatial coupling~\cite{sinha:1992,kozma:1998} has been explored where the number of neighbors included is increased from just the two nearest neighbors. The spatial extent plays a crucial role in the emergent spatiotemporal dynamics of CMLs, such as synchronization, coherence, and chimera states~\cite{omelchenko:2011}. It has also been observed that incorporating long-range spatial interactions enhances synchronization and dampens chaotic dynamics~\cite{wang:2002-synchronization, dos2007lyapunov,batista:2020}. In this work, we use the CLVs to probe fundamental features of spatiotemporal chaos as the extent of the spatial coupling is varied.

The reduced computational complexity of CMLs allows us to probe high-dimensional dynamics using the CLVs, over a wide range of parameters.  We are able to investigate the dynamics for very long times and for many initial conditions when desired. Furthermore, we can specifically tailor the manner in which the maps are spatially coupled with one another. This allows us to quantify the strength and extent of the spatial coupling.  An extensive study, such as this, would be extremely difficult to conduct for a laboratory-scale system such as fluid convection. 

We previously explored the role of nearest-neighbor diffusive coupling on the chaotic dynamics of 1D lattices of coupled-maps~\cite{raj:2024}.  Here, we extend this approach to explore the role of extended spatial coupling, beyond nearest neighbor interactions, on the chaotic dynamics. We develop a theoretical description that predicts the variation of the spectral properties of the CLVs and the values of the Lyapunov exponents as the spatial extent of the coupling is increased. We use these findings to provide new physical insights into the chaotic dynamics of systems with extended spatial coupling.

The remainder of the paper is organized as follows. In \S\ref{section:approach} we discuss our general approach using a 1D lattice of spatially coupled maps. In \S\ref{section:results} we investigate the dynamics for a broad range of spatial couplings and develop theoretical predictions of the Lyapunov exponents and CLVs using the eigenvalues and eigenvectors of the coupling matrix.  The physical insights gained from the CLVs and covariant Lyapunov exponents (CLEs) are discussed. In \S\ref{section:conclusion} our conclusions are presented.

\section{Approach}
\label{section:approach}

\begin{figure}[h!]
\begin{center}
\includegraphics[width=2.75in]{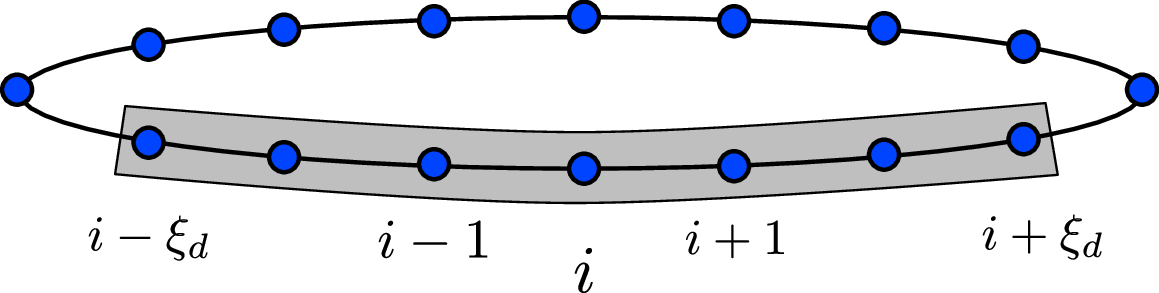}
\end{center}
\caption{A 1D spatially periodic lattice of maps with $N$ sites and extended spatial coupling. The lattice site index is $i$ and the gray region indicates the extent of the spatial coupling about site $i$. For illustration, we show $N\!=\!16$ and $\xi_d \!=\! 3$ for a total of 6 neighbors that are coupled. In our study we use $N\!=\!256$ and $1 \!\le\! \xi_d \!\le\! 12$.}
\label{fig:lattice}
\end{figure}

Our model is the 1D lattice of maps with spatial coupling given by
\begin{equation}
u_i^{(n+1)} = (1 - \epsilon)f(u_i^{(n)}) + \frac{\epsilon}{2 \xi_d} \sum_{j=1}^{\xi_d} \left( f(u_{i+j}^{(n)}) + f(u_{i-j}^{(n)}) \right)
\label{eq:extended-diffusion}
\end{equation}
where $u_i^{(n)}$ is a real value at lattice site $i$ and discrete time $n$. There are $N$ lattice sites on the spatially periodic lattice where $\epsilon$ is the strength of the spatial coupling and $\xi_d$ is the number of neighboring lattice sites, to the left or right of site $i$, that are included in the coupling. We assume a spatially symmetric coupling, however this is not required and an asymmetric coupling could be explored if desired. Figure~\ref{fig:lattice} shows a lattice using our conventions. $\xi_d \!=\! 1$ is the case of nearest neighbor diffusive coupling.

The spatial coupling in Eq.~(\ref{eq:extended-diffusion}) is not meant to represent diffusive coupling for $\xi_d \!>\! 1$. Instead, it is a spatial coupling that involves the average of the dynamics over a group of neighbors whose spatial extent is determined by $\xi_d$. We investigate how increasing the number of coupled neighbors affects fundamental features of the chaotic dynamics.

Chaos is generated locally in Eq.~(\ref{eq:extended-diffusion}) by applying a nonlinear mapping function, $f(u)$, to all values of $u$ at the previous time step. Each lattice site undergoes a nonlinear mapping step to update its value, and then all of the lattice sites are spatially coupled to arrive at the lattice values for the next time step.

The dynamics of the coupled maps can be expressed as
\begin{equation}
    \vec{u}^{(n+1)} = \mathbf{A}_c(\epsilon,\xi_d) \vec{f}(\vec{u}^{(n)})
\end{equation}
where $\vec{u}$ is the $N$-dimensional vector of the values of $u_i$ for all $i$. $\mathbf{A}_c(\epsilon,\xi_d)$ is the banded $N \!\times\! N$ spatial coupling matrix with $2 \xi_d \!+\! 1$ bands. In our exploration, $2 \xi_d \!+\! 1 \!\ll\! N$ and $\mathbf{A}_c$ remains sparse for all of the conditions we investigate. $\mathbf{A}_c$ depends only upon the details of the spatial coupling, and it is independent of the mapping function $f(u)$.

We use the centered quadratic map as the mapping function 
\begin{equation}
f(u) = r \left( \frac{1}{4} - u^2 \right)
\label{eq:quadratic-map}
\end{equation}
where $r$ is the control parameter.  The parameter space that describes the dynamics of the CML, given by $\{r,\epsilon,\xi_d,N\}$, is vast and it is not our intention to provide an exhaustive analysis of the rich dynamics over this space. Instead, we focus on investigating the role of extended spatial coupling on a lattice that is exhibiting chaotic dynamics.

In this light, we choose $r\!=\!2.8, \epsilon\!=\!0.7, N\!=\!256$ and vary $\xi_d$ over the range $1 \!\le\! \xi_d \!\le\! 12$. These values have been chosen for several reasons. For these values of $\{r,\epsilon,N\}$ the dynamics have been investigated in detail for $\xi_d\!=\!1$~\cite{raj:2024} which provides an insightful baseline of understanding. For these conditions, the dynamics were shown to be extensively chaotic.  As a result, it is expected that the chosen system size $N$ is sufficient to be in the large system limit where statistical properties of the chaos, such as the dimension density, would not change significantly for larger systems. In addition, it was found that the leading Lyapunov exponent remained approximately constant over a wide range of diffusion strengths $0.2 \!\lesssim\! \epsilon \!\lesssim\! 1$. A single isolated quadratic map undergoes a period doubling route to chaos as $r$ is increased. A value of $r \!=\! 2.8$ leads to chaotic dynamics with a Lyapunov exponent of $\lambda_0 \!=\! 0.6034$.  Overall, these parameters are chosen to generate chaotic dynamics which we explore in detail as the spatial extent of the coupling is increased. We have not explored how the chaotic dynamics vary as either the strength of the diffusion $\epsilon$ or the control parameter $r$ are varied. The dynamics of the system is very rich, and it would be interesting to pursue the influence of these parameters in future work. 

\section{Results}
\label{section:results}

\subsection{Spatiotemporal Dynamics}
\label{section:spatiotemporal-dynamics}

\begin{figure}[h!]
\begin{center}
\includegraphics[width=2.5in]{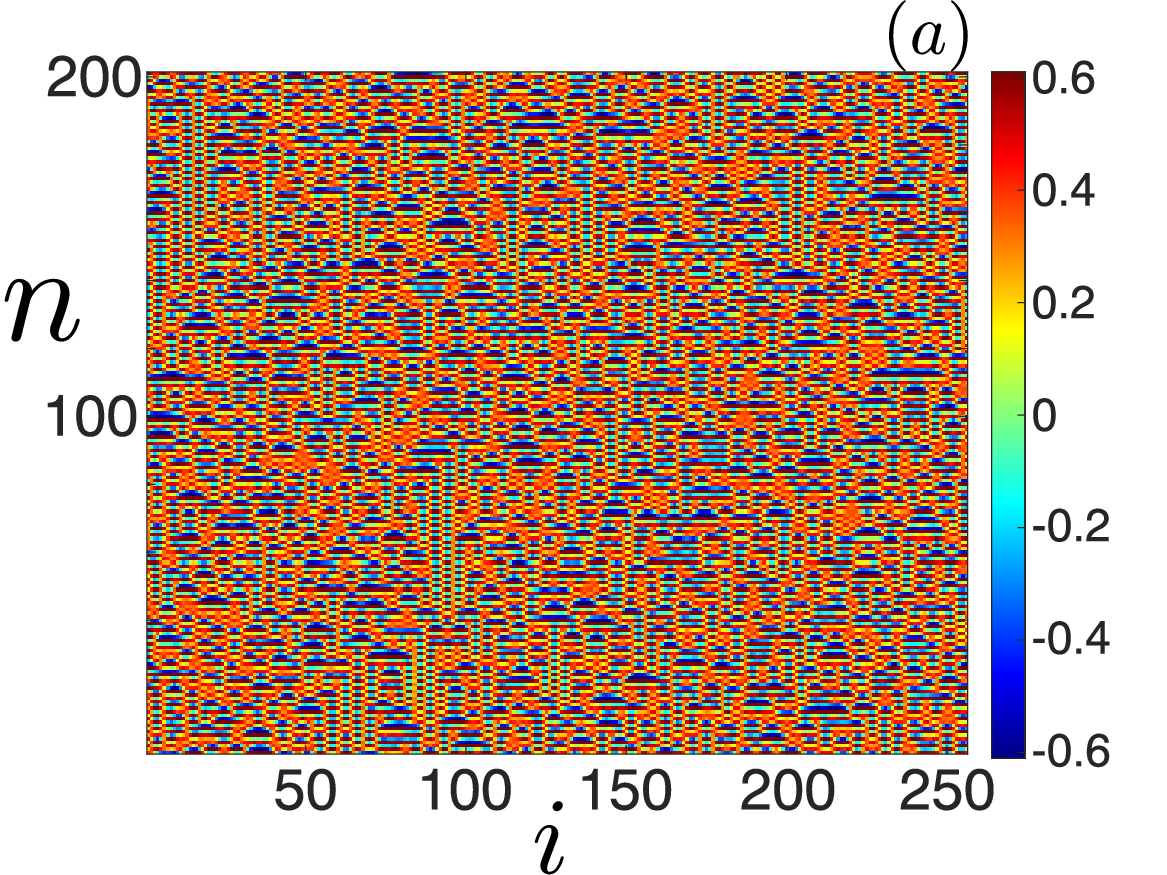} \hspace{0.4cm}
\includegraphics[width=2.5in]{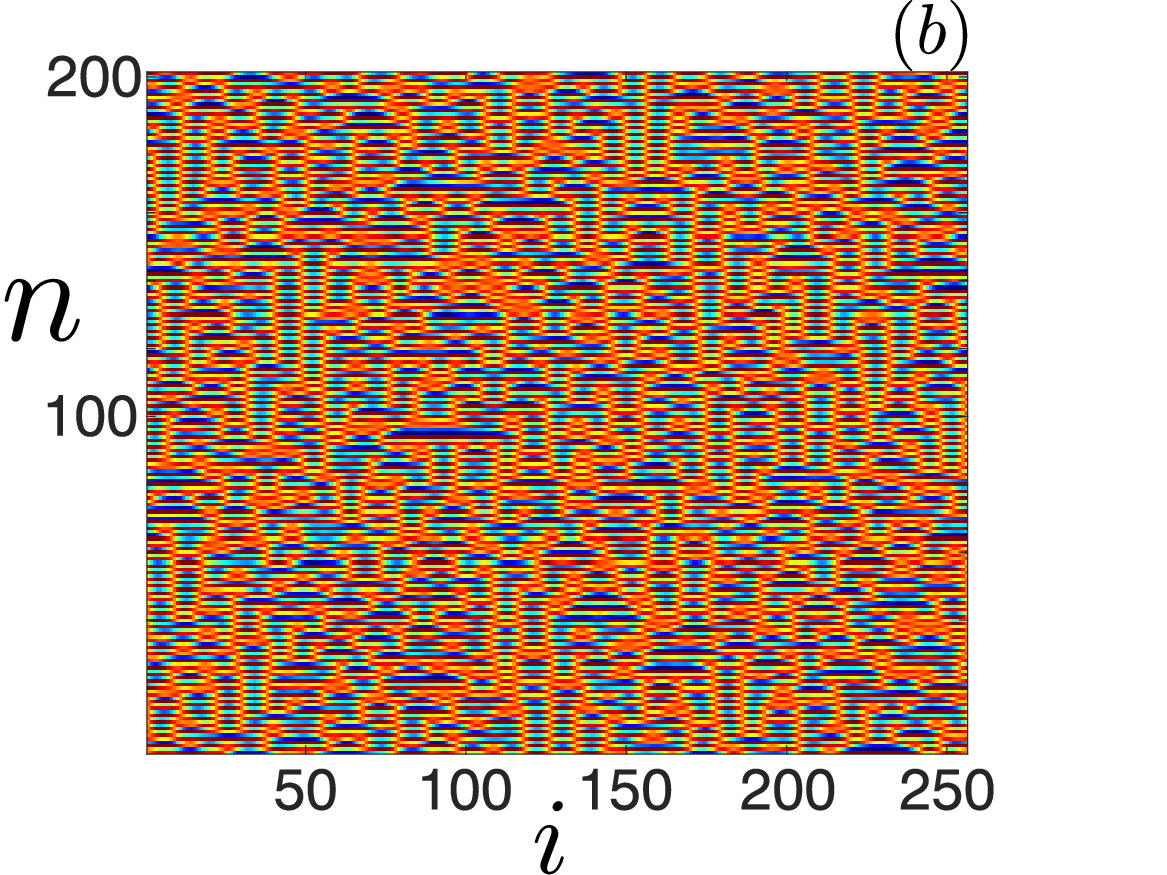} \\ 
\includegraphics[width=2.5in]{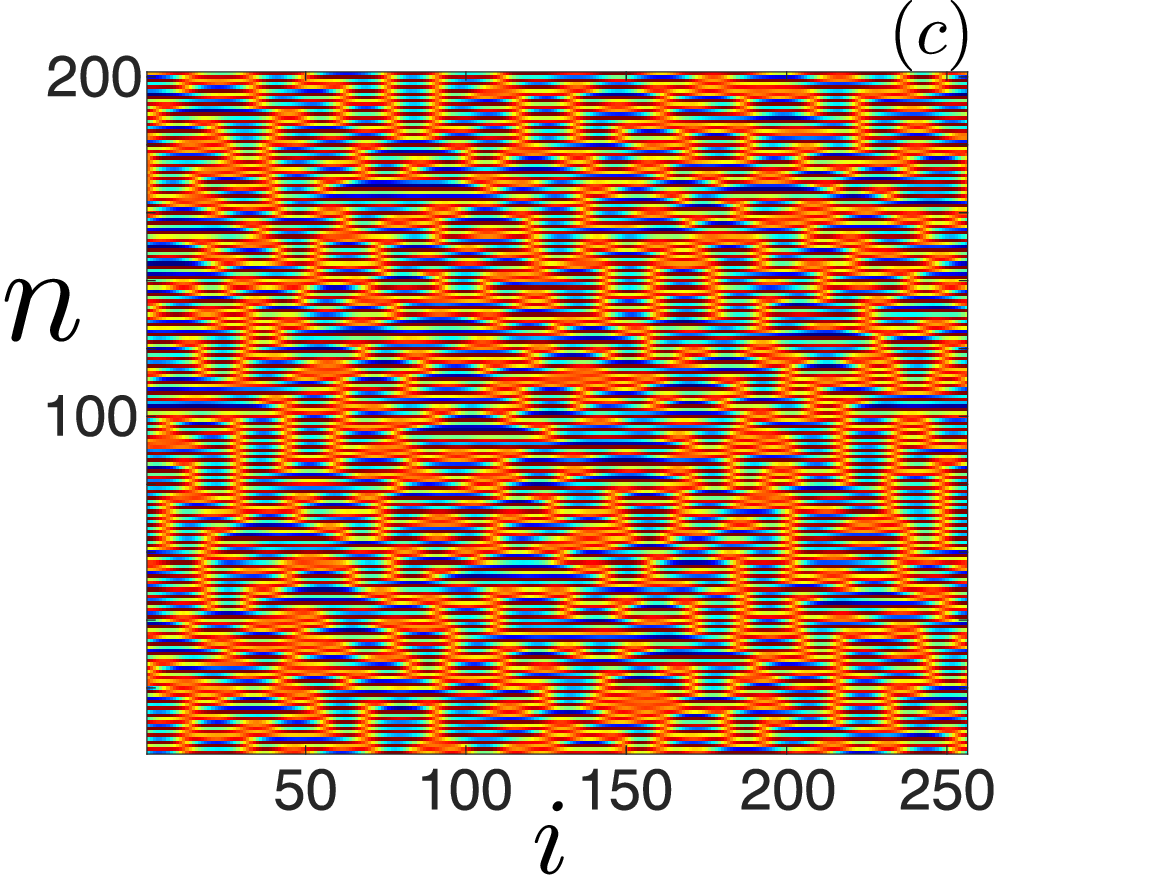} \hspace{0.4cm}
\includegraphics[width=2.5in]{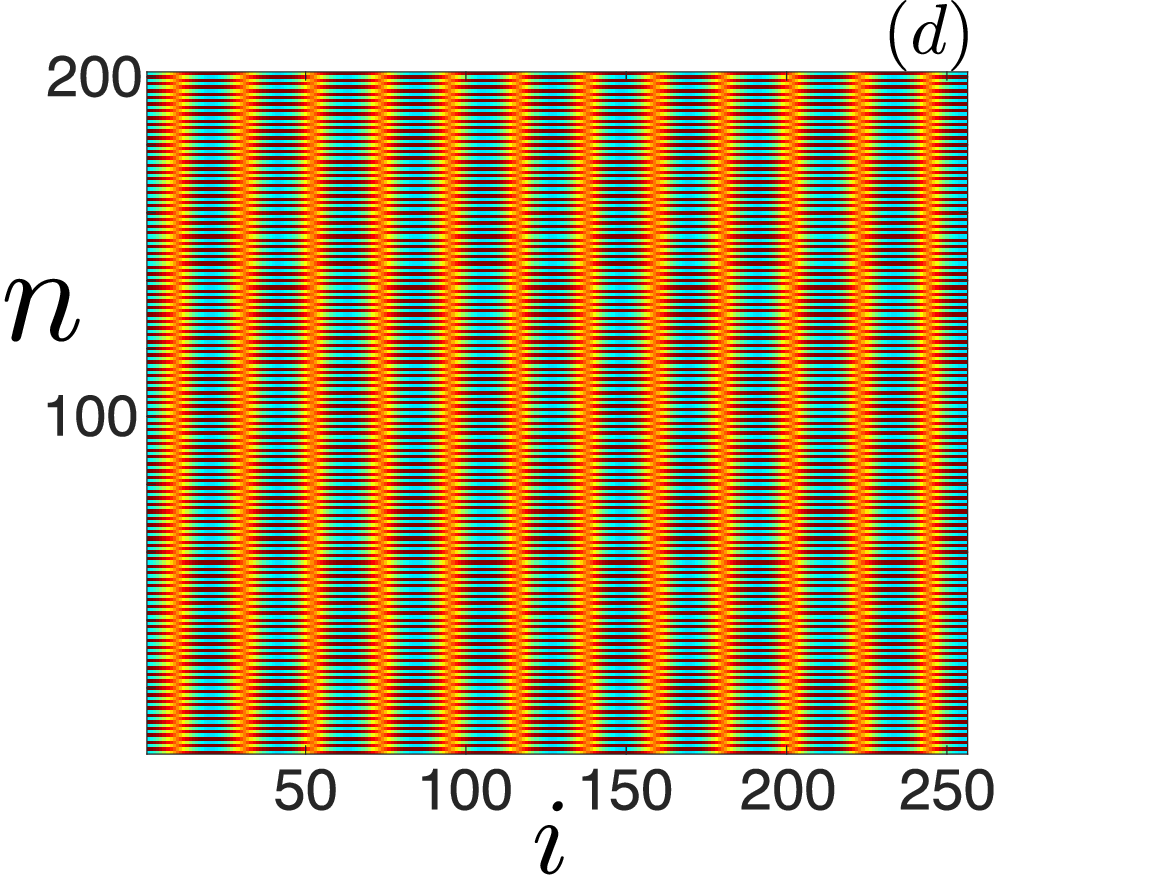}
\end{center}
\caption{Spacetime plots of $u_i^{(n)}$: ($a$)~$\xi_d \!=\! 1$ (nearest-neighbor coupling), ($b$)~$\xi_d \!=\! 3$ (6-neighbor coupling), ($c$)~$\xi_d \!=\! 6$ (12-neighbor coupling), ($d$)~$\xi_d \!=\! 10$ (20-neighbor coupling).}
\label{fig:extended-diff-spacetime}
\end{figure}

The lattice dynamics are computed by iterating Eq.~(\ref{eq:extended-diffusion}) forward in time from random initial conditions. Spacetime plots of the dynamics are shown in Fig.~\ref{fig:extended-diff-spacetime} for four values of $\xi_d$. Figure~\ref{fig:extended-diff-spacetime} shows the lattice values for two hundred time steps after an initial $2 \times 10^6$ time steps have been completed to allow initial transients to decay. The color contours indicate the value of $u_i^{(n)}$, all panels use the color bar shown in~Fig.~\ref{fig:extended-diff-spacetime}($a$).

Figure~\ref{fig:extended-diff-spacetime}($a$) shows chaotic dynamics for nearest-neighbor diffusive coupling ($\xi_d\!=\!1$). The chaotic dynamics include spatial structures that exceed the lattice spacing length scale. Figure~\ref{fig:extended-diff-spacetime}($b$)-($c$) illustrate how the chaotic dynamics change as the length of the spatial coupling increases from 6 to 12 neighbors, respectively. It is evident that the length scale of the spatial structures in the dynamics increases with increasing $\xi_d$. Figure~\ref{fig:extended-diff-spacetime}($d$) illustrates periodic dynamics for the case of $\xi_d\!=\!10$.

\begin{figure}[h!]
\begin{center}
\includegraphics[width=2.5in]{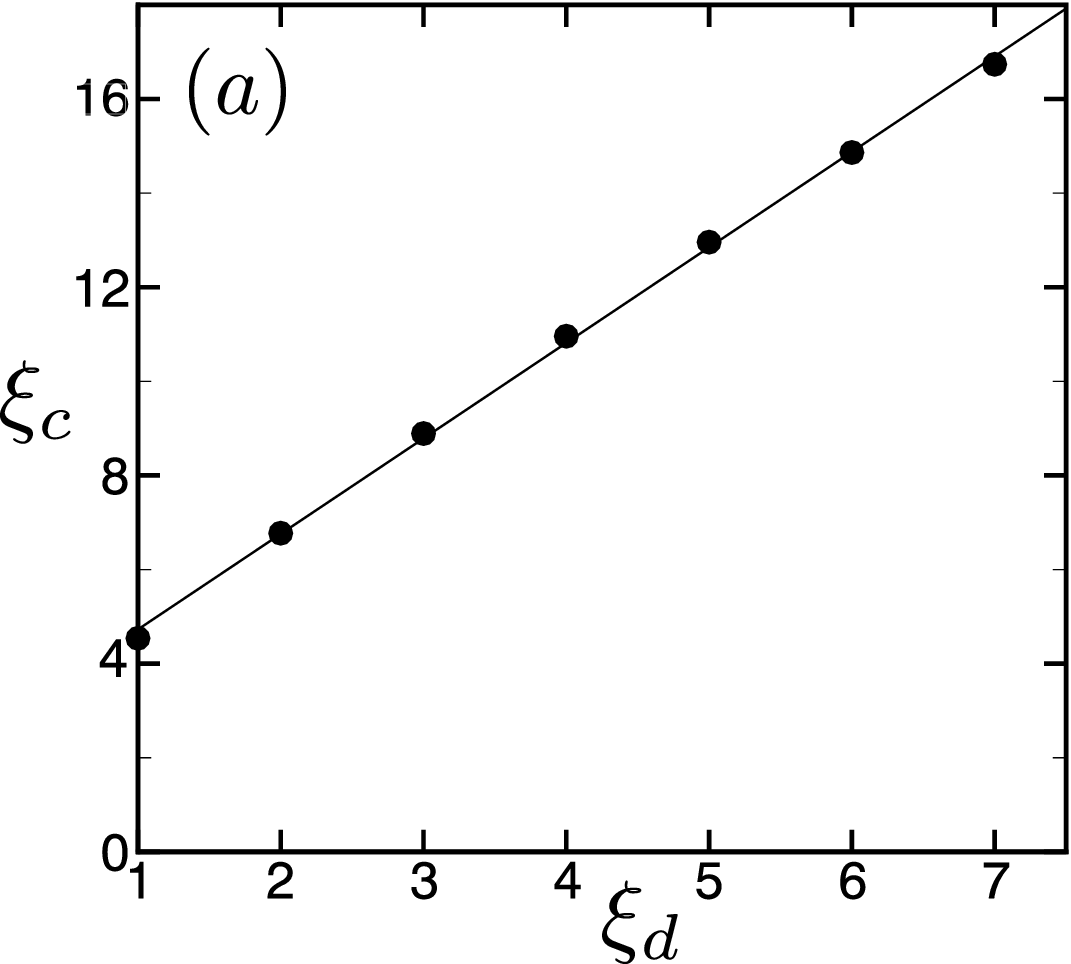} \hspace{0.4cm}
\includegraphics[width=2.5in]{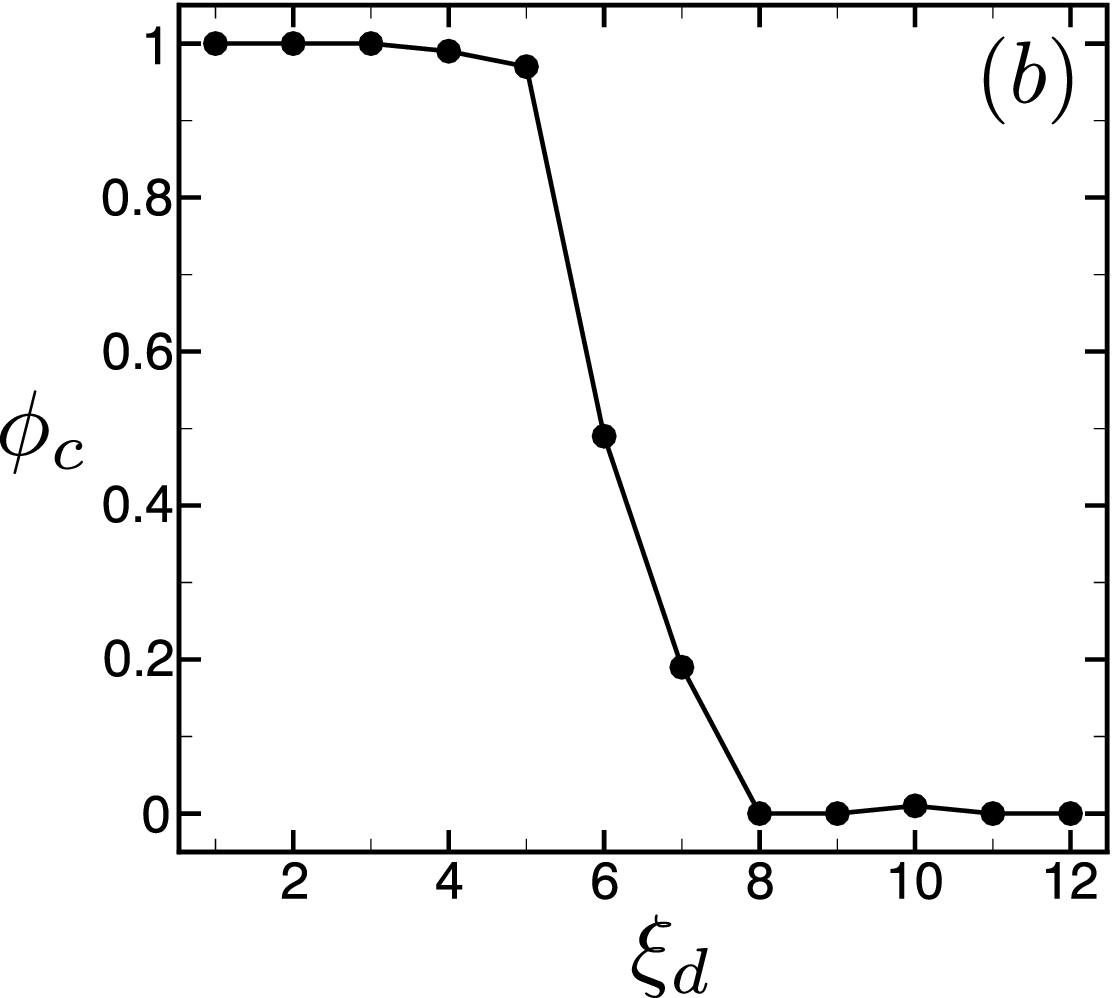}
\end{center}
\caption{($a$) The linear variation of the correlation length $\xi_c$ with the length scale of the spatial coupling $\xi_d$. The solid line is the curve-fit $\xi_c \!=\! 2.70 \!+\! 2.03 \xi_d$.  ($b$) The fraction $\phi_c$ of random initial conditions which lead to chaotic dynamics. Each simulation is iterated forward in time for $2 \times 10^6$ time steps and 100 simulations were conducted for each value of $\xi_d$ using different random initial conditions.}
\label{fig:xic}
\end{figure}

To quantify the length scale of the structures shown in Fig.~\ref{fig:extended-diff-spacetime}, the two-point correlation length $\xi_c$ has been computed. The value of $\xi_c$ is determined as the first zero crossing of the two-point correlation.  The variation of $\xi_c$ with $\xi_d$ is shown in Fig.~\ref{fig:xic}($a$) where $\xi_c$ is only calculated for values of $\xi_d$ yielding chaotic dynamics. The solid line is the curve-fit $\xi_c \!=\! 2.70 \!+\! 2.03 \xi_d$ yielding a linear growth spatial correlations with $\xi_d$ as expected. A unit change of $\xi_d$ includes two additional neighbors in the coupling which is reflected by the slope of $\sim \! 2$ in Fig.~\ref{fig:xic}($a$).

As $\xi_d$ is increased, there is a transition from chaotic to periodic dynamics.  In Fig.~3($b$), we show the fraction of random initial conditions, $\phi_c$, that yield a lattice with chaotic dynamics as indicated by a positive leading Lyapunov exponent $\lambda_1 >0$. Each data symbol is the fraction that has been computed using 100 different random initial conditions for that value of $\xi_d$ where each individual simulation has been evolved forward in time for $2 \times 10^6$ time units.  For $\xi_d \!\le\! 5$ nearly all initial conditions lead to chaotic dynamics $\phi_c \!\approx\! 1$. This is followed by a rapid falloff to nearly all periodic dynamics for $\xi_d \!\ge\! 8$ where $\phi_c \!\approx\! 0$ (for example, see  Fig.~\ref{fig:extended-diff-spacetime}($d$)). We have not explored the coexistence of periodic and chaotic solutions further which occurs for $\xi_d \!=\! \{6,7\}$. In the following, we focus our investigation upon system parameters that yield chaotic dynamics.

\subsection{Chaotic Dynamics}
\label{section:dynamics}

We next consider how increasing the length scale of the spatial coupling affects the dynamics as indicated by the CLVs. The CLVs describe the growth, or decay, of small perturbations to the dynamics along the nonlinear trajectory in the state-space given by the evolution of Eq.~(\ref{eq:extended-diffusion})~\cite{pikovsky:2016}.  There are a total of $N$ CLVs, $\vec{v}_k$, where $k \!=\! 1,2,\ldots,N$ is the Lyapunov index and each CLV is a $N$-dimensional vector.  The numerical approach for computing the CLVs~\cite{ginelli:2007} using our conventions is described in detail in Ref.~\cite{raj:2024}.

The essential steps for computing the CLVs are the following.  $N$ copies of the linearized tangent-space equations are simultaneously computed along with Eq.~(\ref{eq:extended-diffusion}) to calculate $N$ orthonormal Lyapunov vectors OLVs (often called the Gram-Schmidt Lyapunov vectors~\cite{wolf:1985}). The OLVs are then evolved backwards in time in order to construct the CLVs as a linear combination of the OLVs using the dynamic algorithm of~\citet{ginelli:2007}.

The CLVs are covariant with the dynamics in the sense that they are oriented in physically meaningful directions in the tangent space, and their magnitude will grow (or decay) according to the value of their  Lyapunov exponent in both forward and backward evolution in time.  The final result is $N$ CLVs which are ordered in descending order by the value of their Lyapunov exponent.

The CLEs are the Lyapunov exponents that are computed using the CLVs. The finite-time CLEs do not equal the finite-time Lyapunov exponents computed using the OLVs. However, the CLEs and the Lyapunov exponents computed using the OLVs are equal to each other in the infinite time limit (long-time limit in practice)~\cite{takeuchi:2011}. The values of the CLEs in the long-time limit will be referred to as the Lyapunov exponents $\lambda_k$ where $k = 1,2, \ldots, N$ is the Lyapunov index.

The numerical procedure for computing the CLVs and CLEs is the following. Equation~(\ref{eq:extended-diffusion}) is iterated forward in time for $2 \! \times \! 10^6$ time steps from random initial conditions. We next compute $N$ OLVs by simultaneously evolving Eq.~(\ref{eq:extended-diffusion}) and $N$ copies of the tangent-space equations forward in time for $8 \!\times\! 10^4$ time steps. During this evolution, the perturbation vectors are reorthogonalized after every time step using a QR decomposition and we save the $\mathbf{Q}$ and $\mathbf{R}$ matrices for later use in the dynamic algorithm. The OLVs and the dynamics are next iterated backwards in time for $6 \!\times\! 10^4$ time units. During this backwards evolution, the converged CLVs are computed as a linear combination of the OLVs. The converged CLVs are then iterated forward in time for $5 \!\times\! 10^4$ time units where the finite time CLEs are computed after every time step. The long-time average of the finite-time CLEs yields the spectrum of Lyapunov exponents $\lambda_k$. The CLVs and CLEs that are computed during this forward time evolution are used in our analysis.

\begin{figure}[h!]
\begin{center}
\includegraphics[width=2.75in]{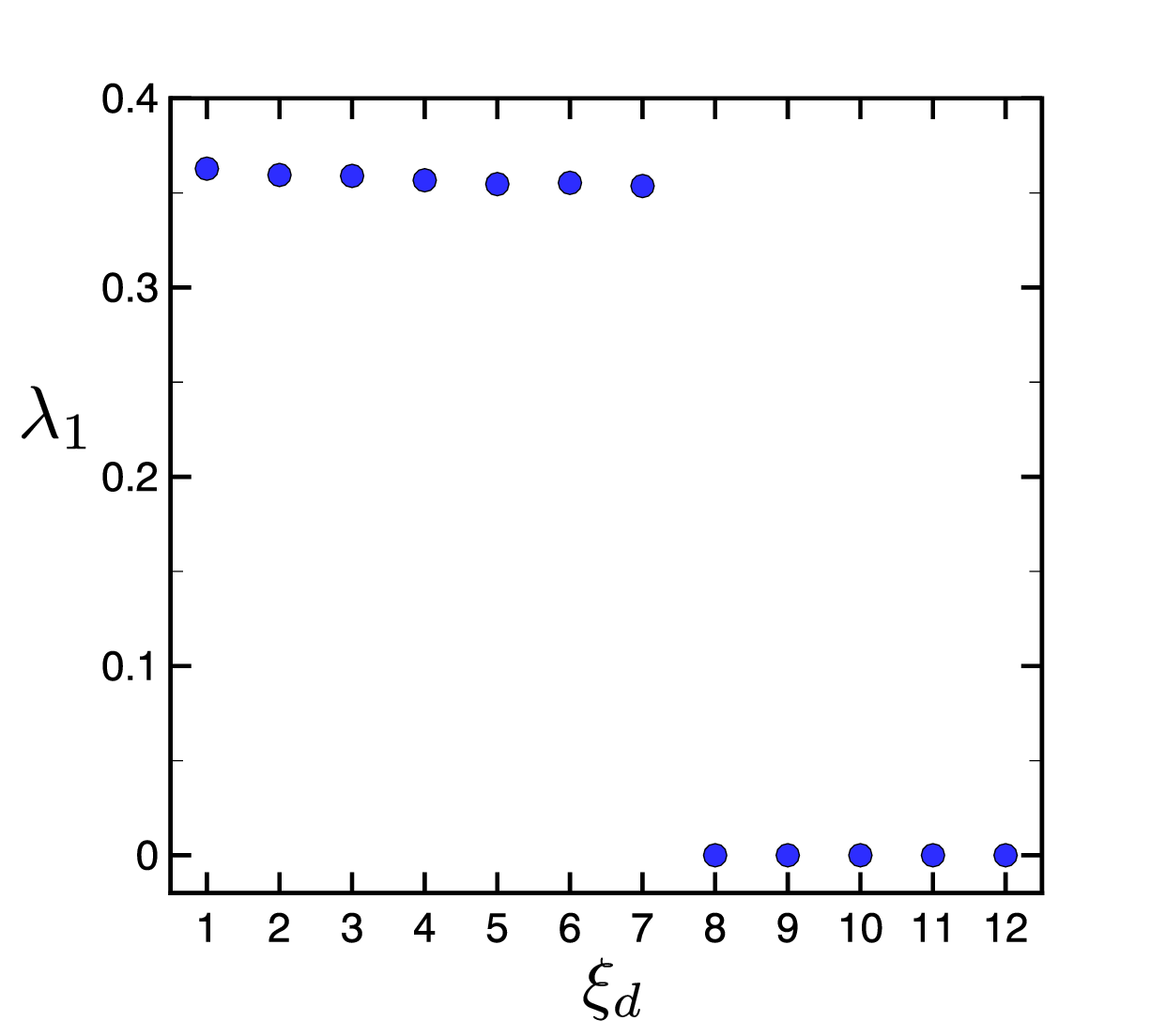}
\end{center}
\caption{The variation of the leading Lyapunov exponent $\lambda_1$ with the extent of the spatial coupling $\xi_d$.  $\lambda_1$ is not affected significantly as the coupling increases to include 14 neighbors ($\xi_d \!=\! 7$). For larger values, $\xi_d \!\ge\! 8$, the dynamics are periodic in time as indicated by the vanishing values of $\lambda_1$.}
\label{fig:lyap-with-xid}
\end{figure}

Figure~\ref{fig:lyap-with-xid} shows the variation of the leading Lyapunov exponent  $\lambda_1$ with $\xi_d$. The lattice dynamics are chaotic for all spatial couplings $\xi_d \!\le\! 7$ with $\lambda_1 \!\approx\! 0.35$. For $\xi_d \!\ge\! 8$ the dynamics are periodic which yields $\lambda_1 \approx 0$. These results indicate that the inclusion of additional neighbors in the spatial coupling does not significantly affect $\lambda_1$ when the dynamics are chaotic.  The value of $\lambda_1$ depends upon the generation of local disorder through the choice of the mapping function and its parameters such as $r$ for the quadratic map of Eq.~\eqref{eq:quadratic-map}.  The spatial coupling present in the lattice yields a reduction in the leading Lyapunov exponent such that $\lambda_1 \!<\! \lambda_0$.

\subsection{Spectra of Lyapunov Exponents}
\label{section:lyapunov-spectra}

\begin{figure}[h!]
\begin{center}
\includegraphics[width=2.75in]{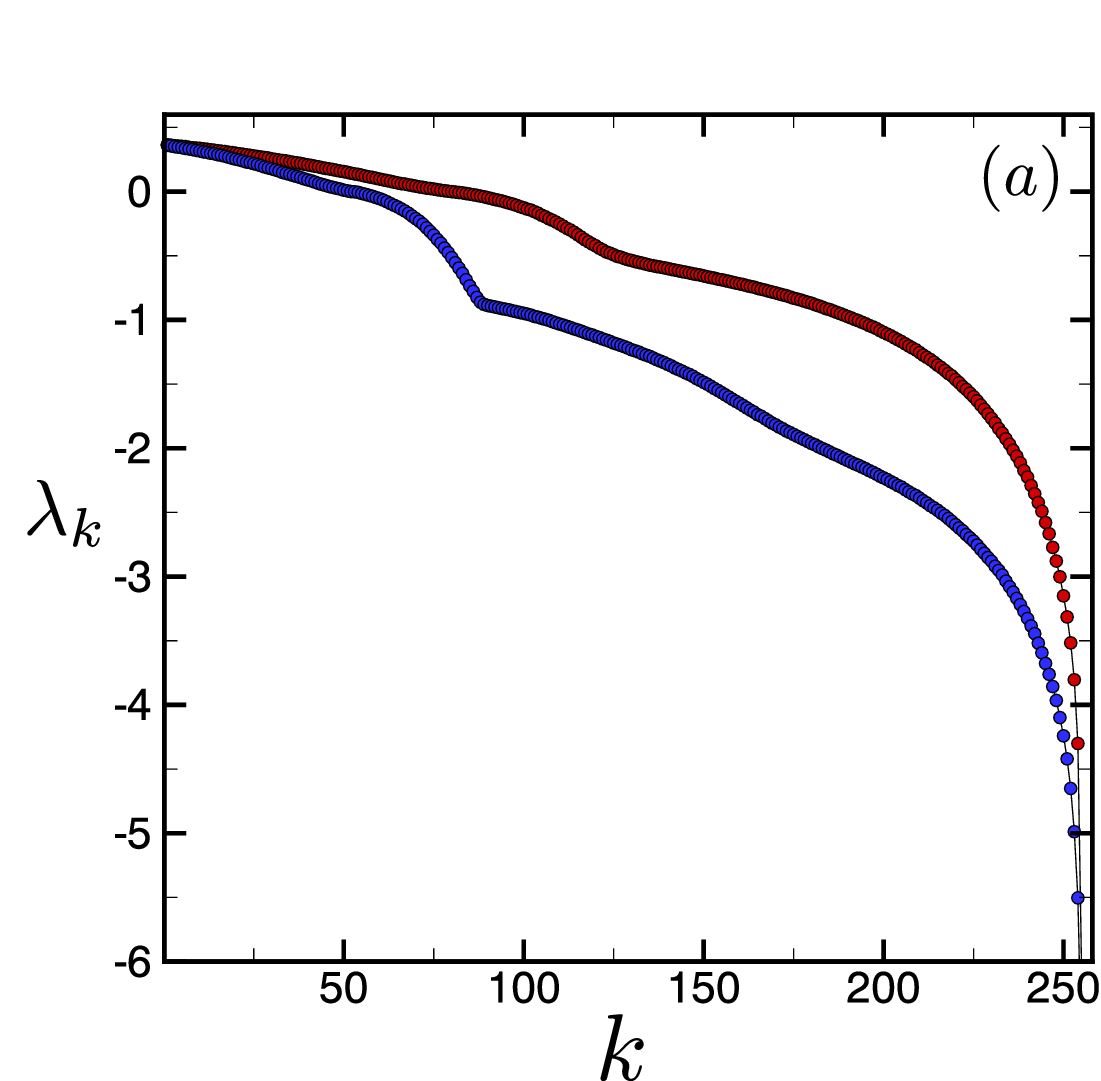} \hspace{0.4cm}
\includegraphics[width=2.75in]{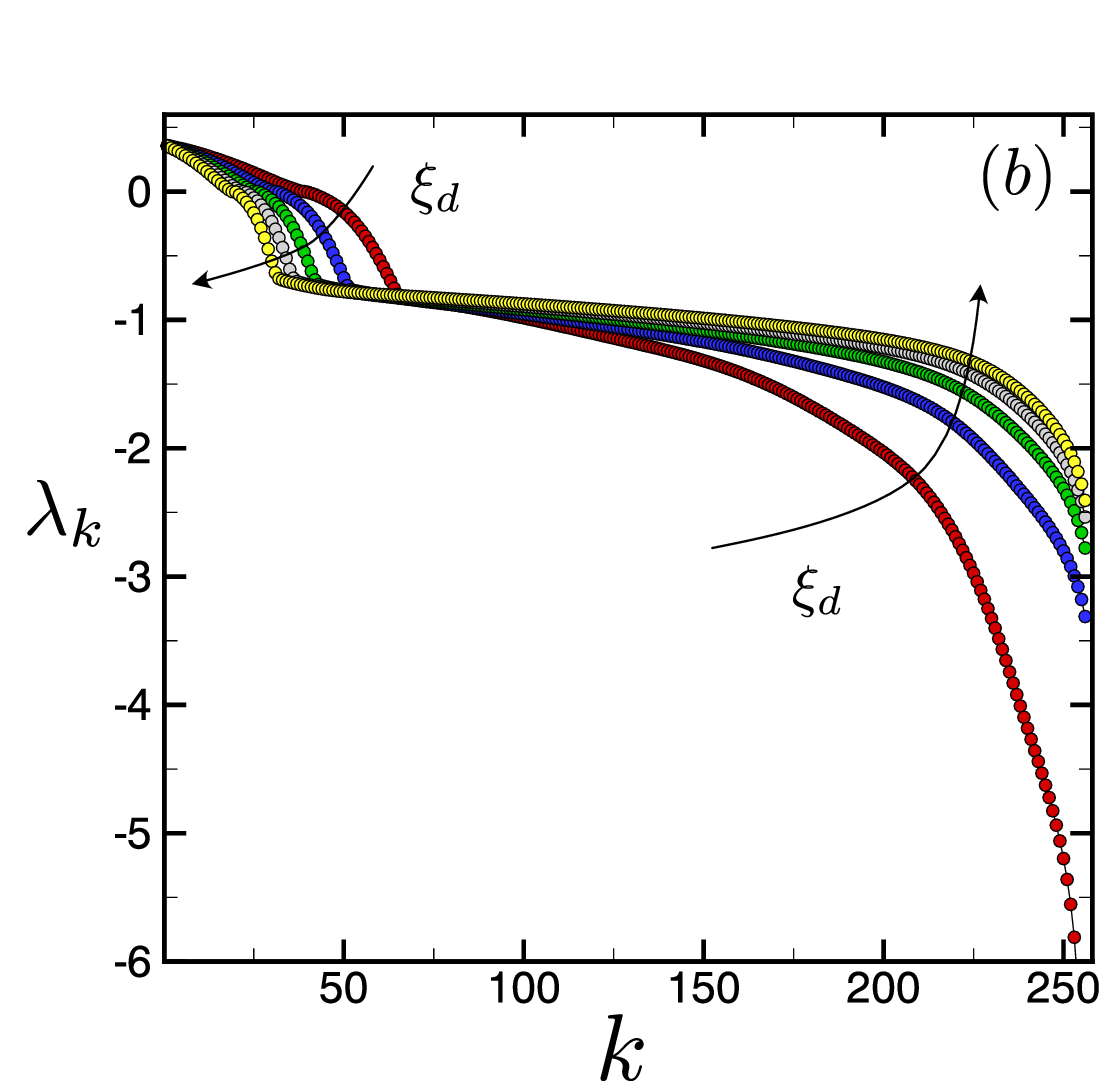}
\end{center}
\caption{Variation of the spectrum of Lyapunov exponents, $\lambda_k$, with the length scale of spatial coupling,  $\xi_d$, where $k$ is the Lyapunov index. ($a$) $\xi_d \!=\! 1$ (upper curve, red), $\xi_d \!=\! 2$ (lower curve, blue). ($b$) Results for $3 \!\le\! \xi_d \!\le\! 7$. The direction of increasing $\xi_d$ is indicated by the arrows to highlight to opposing trends found for small and large $k$.}
\label{fig:lyapunov-spectrum}
\end{figure}

The spectra of Lyapunov exponents $\lambda_k$ is shown in Fig.~\ref{fig:lyapunov-spectrum}. Results for nearest neighbor coupling $\xi_d\!=\!1$ (red, upper) and for coupling with four neighbors $\xi_d\!=\!2$ (blue, lower) are shown in Fig.~\ref{fig:lyapunov-spectrum}($a$).  The leading Lyapunov exponents for these cases are nearly identical, as indicated by Fig.~\ref{fig:lyap-with-xid}, and both spectra decay with increasing values of the Lyapunov index $k$. Overall, the decay of the Lyapunov spectrum for $\xi_d\!=\!2$ is larger than for the case of nearest neighbor coupling. In addition, the spectra contain interesting elbow features at $k \!\approx\! 128$ for $\xi_d\!=\!1$ and $k \!=\! 88$ for $\xi_d\!=\!2$.

Figure~\ref{fig:lyapunov-spectrum}($b$) shows the variation of the Lyapunov spectrum over a range of spatial couplings where $3 \!\le\! \xi_d \!\le\! 7$. The Lyapunov spectra show two opposing trends. For small Lyapunov index $k$, the value of $\lambda_k$ decreases with increasing spatial coupling $\xi_d$, as indicated by the arrow in the upper left portion of Fig.~\ref{fig:lyapunov-spectrum}($b$). We emphasize that this includes all of the positive Lyapunov exponents and some of the negative Lyapunov exponents. However, for the larger negative Lyapunov exponents, at larger $k$, the Lyapunov exponents increase with increasing values of $\xi_d$ as indicated by the arrow on the right side of Fig.~\ref{fig:lyapunov-spectrum}($b$). The directions of these two opposing trends are indicated by the arrows which point in the direction of the spectra variation with increasing $\xi_d$.

Figure~\ref{fig:lyapunov-spectrum}($b$) indicates that the values of the positive Lyapunov exponents (excluding $\lambda_1$), and a portion of the small negative exponents which occur before the elbow, decrease with increasing spatial coupling. At the same time, the values of the large negative Lyapunov exponents increase with increasing coupling. 

Insight into the Lyapunov spectra can be gained from the eigenvalues, $\Lambda_k$, and eigenvectors, $\vec{w}_k$, of the spatial coupling matrix $\textbf{A}_c$ where
\begin{equation}
\textbf{A}_c \vec{w}_k = \Lambda_k \vec{w}_k
\label{eq:eigen}
\end{equation}
and $k$ is the eigenvector index. The connection between the Lyapunov exponents and the eigenvalues can be expressed as
\begin{equation}
\lambda_k \!=\! \lambda_1 \!+\! \ln | \Lambda_k|
\label{eq:lyapunov-spectrum}
\end{equation}
where $\Lambda_k$ has been sorted such that the $|\Lambda_k|$ are in descending order for $k = 1,2, \ldots, N$. We emphasize that $\lambda_1$ must be computed numerically for the coupled lattice. There is not currently a way to theoretically predict $\lambda_1$. However, we note that $\lambda_1 \!<\! \lambda_0$ for the spatial coupling we consider and, as a result, $\lambda_0$ could be used to provide an overpredicted estimate. This approach was used by \citet{takeuchi:2011} to describe coupled tent maps and also used to explore the role of nearest neighbor coupling for coupled quadratic maps for a wide range of conditions~\cite{raj:2024}.  

For periodic boundary conditions, $\mathbf{A}_c$ is a circulant matrix~\cite{gray:2006} and the unsorted eigenvalues $\Lambda_{k'}$ are given by 
\begin{equation}
\Lambda_{k'} = 1 - \epsilon \left[ 1 - \frac{1}{\xi_d} \sum_{\xi = 1}^{\xi_d} \cos \left( \frac{2 \pi (k' - 1) \xi}{N} \right) \right].
\label{eq:eigenvalues-unsorted}
\end{equation}
We emphasize that $\Lambda_{k'}$ are the unsorted eigenvalues where the index is $k' \!=\! 1, 2, \!\ldots\!, N$. A theoretical prediction for the Lyapunov spectrum can be found by computing the eigenvalues using Eq.~(\ref{eq:eigenvalues-unsorted}), sorting them by their magnitude, and then using Eq.~(\ref{eq:lyapunov-spectrum}).

\begin{figure}[h!]
\begin{center}
\includegraphics[width=2.75in]{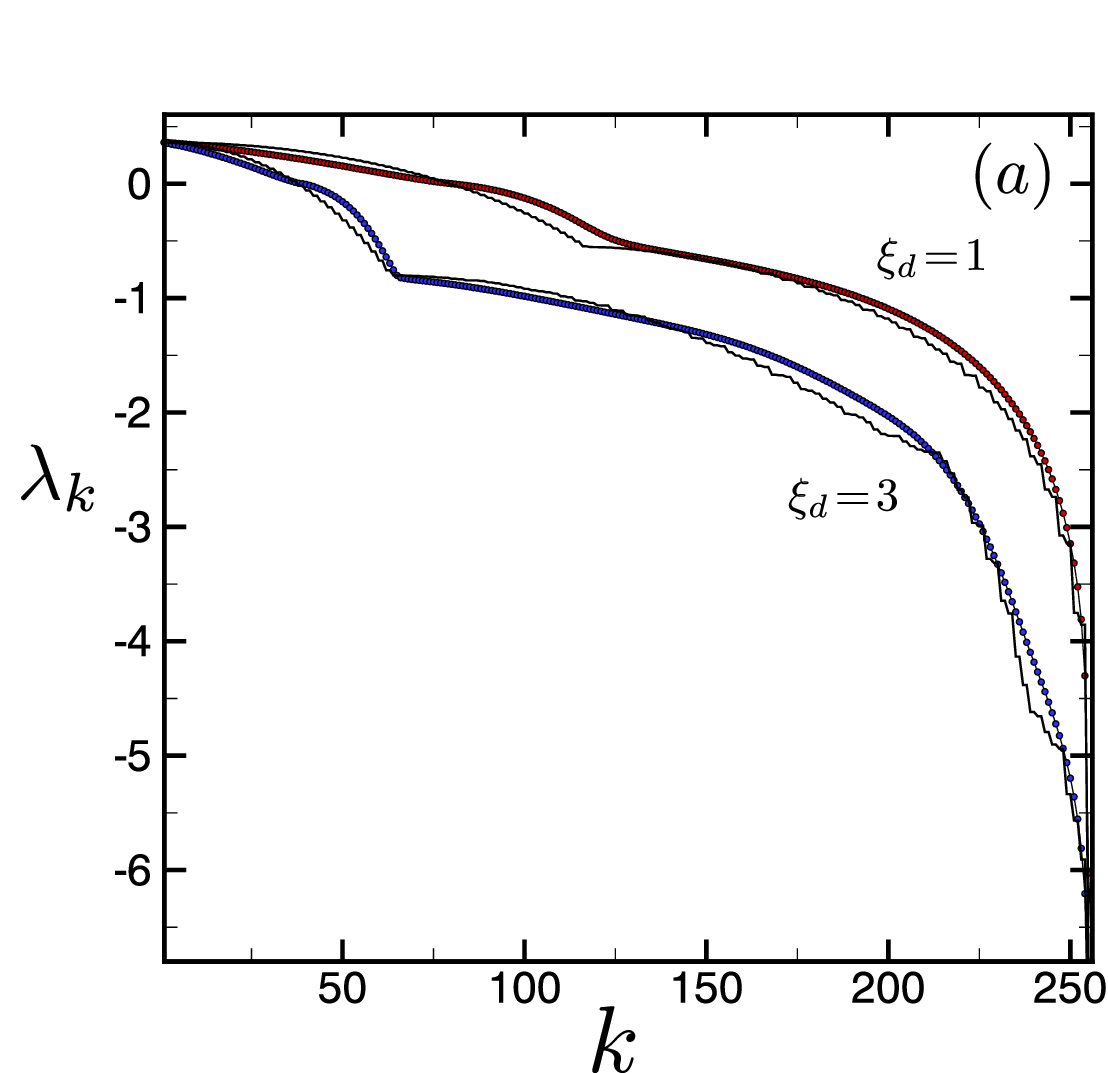} \hspace{0.4cm}
\includegraphics[width=2.75in]{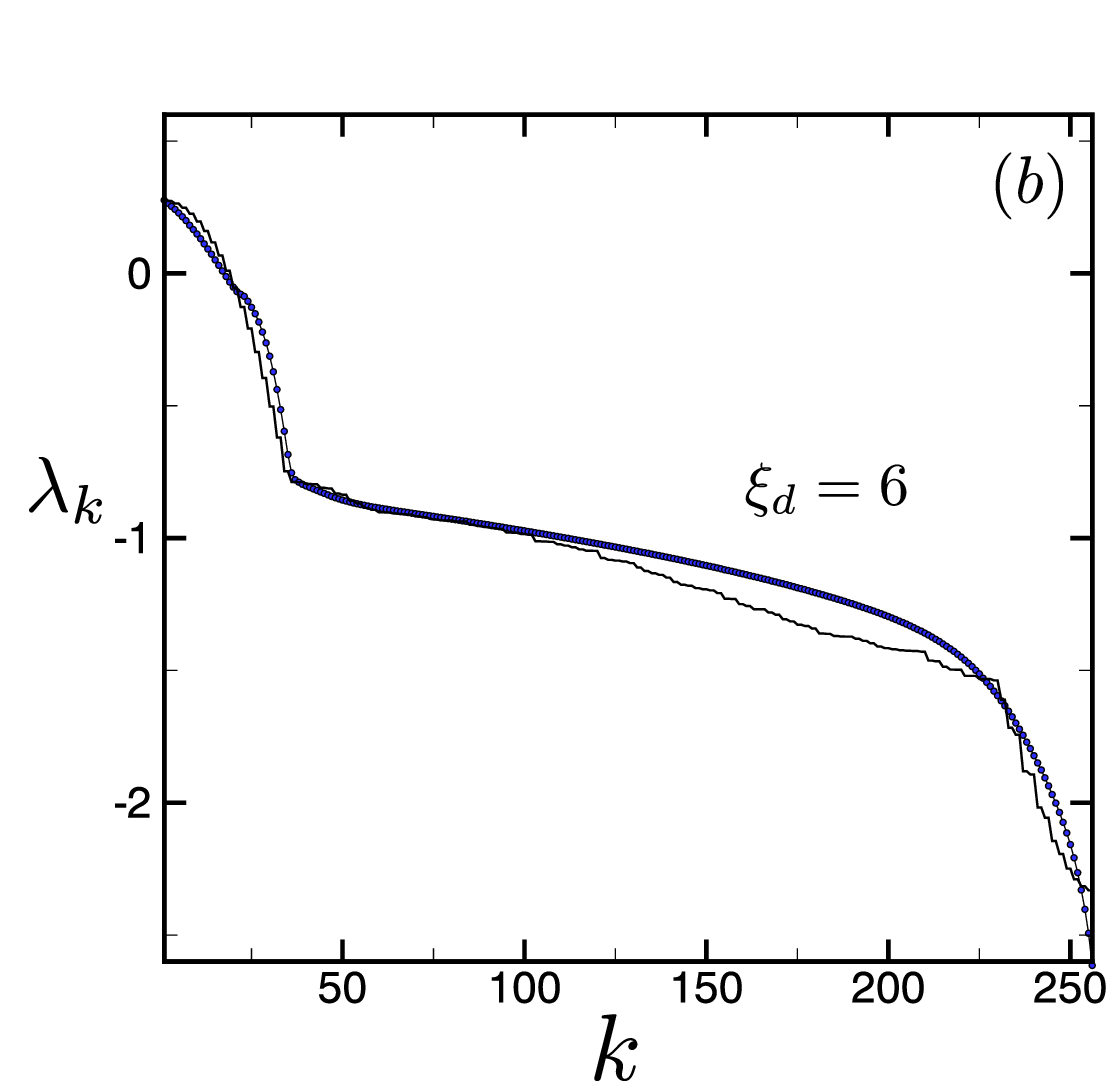}
\end{center}
\caption{A comparison of the computed Lyapunov spectra (symbols) with the theoretical prediction using Eqs.~(\ref{eq:lyapunov-spectrum})-(\ref{eq:eigenvalues-unsorted}) (solid black lines): ($a$) $\xi_d\!=\!1$, $\xi_d\!=\!3$; ($b$) $\xi_d\!=\!6$.}
\label{fig:lyapunov-spectrum-comparison}
\end{figure}

A comparison of the numerically computed Lyapunov spectrum with the theoretical prediction is shown in Fig.~\ref{fig:lyapunov-spectrum-comparison} for three values of $\xi_d$. Overall, the agreement is very good. The addition of more neighbors in the spatial coupling does not effect $\lambda_1$ when the dynamics are chaotic. However, the addition of more neighbors affects the shape of the Lyapunov spectrum which is determined by the eigenvalues of $\mathbf{A}_c$.  We emphasize that this approach requires a spatial coupling that depends linearly upon the values at the different lattice sites which can be represented as a coupling matrix. An important example of a situation where this approach would not work is in the presence of a nonlinear convective coupling.

The fractal dimension $D_\lambda$ (or Kaplan-Yorke dimension) can be calculated from of the Lyapunov exponents~\cite{kaplan:1979} as
\begin{equation}
D_\lambda = p + \frac{\sum_{k=1}^p \lambda_k}{|\lambda_{p+1}|}
\label{eq:fractal-dimension}
\end{equation}
where $p$ is the largest value of the Lyapunov index such that the summation is positive.  The fractal dimension provides an estimate for the number of chaotic degrees of freedom that are active, on average~\cite{farmer:1983}.

\begin{figure}[h!]
\begin{center}
\includegraphics[width=2.75in]{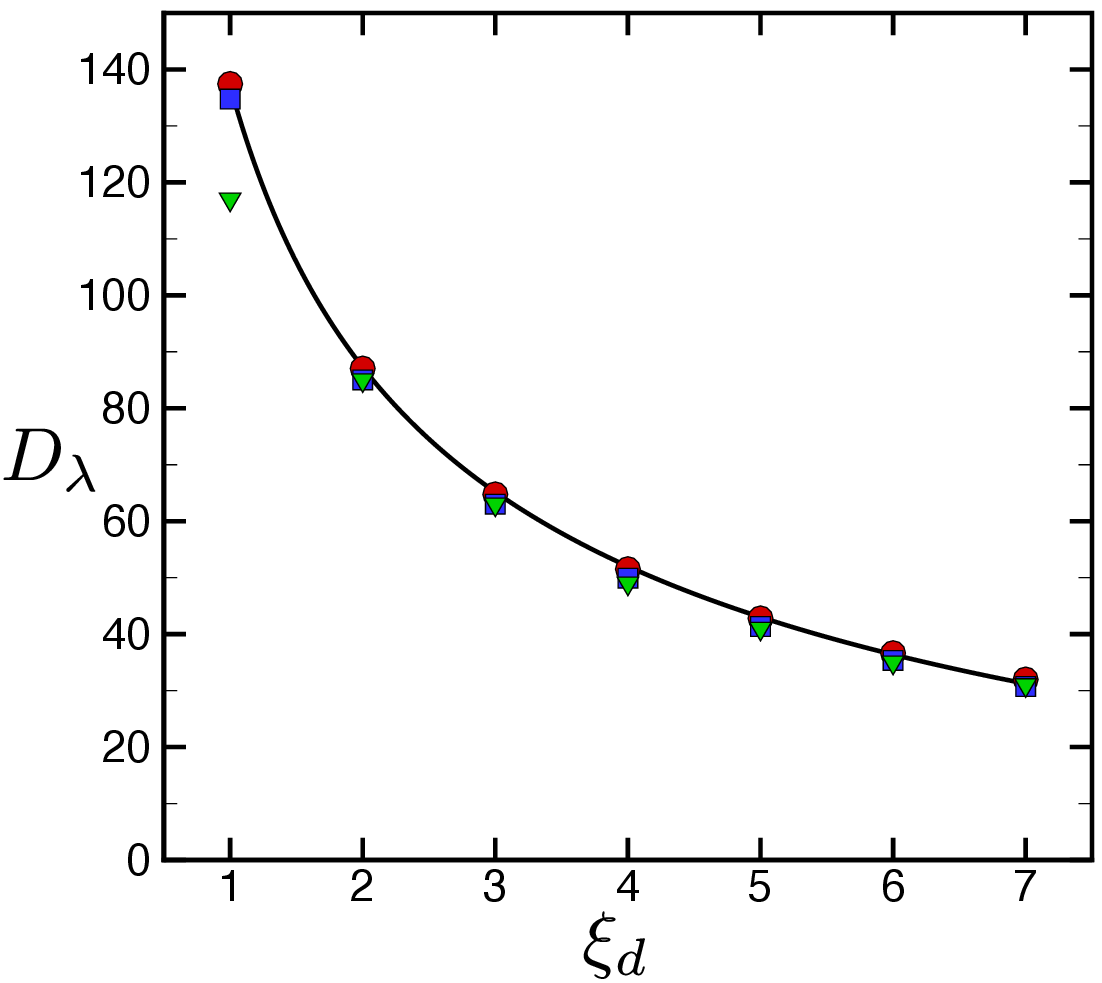}
\end{center}
\caption{Variation of the fractal dimension $D_\lambda$ with $\xi_d$.  Theoretical predictions (squares, blue), numerically computed values (circles, red), and the fit $D_\lambda \!=\! 170.17 \xi_d^{-1/2} \!-\! 33.08$ using the numerical results (solid line). The green triangles are the largest value of the eigenvector index $k^*$ in the linear regime of the spatial power spectrum.}
\label{fig:fractal-dimension}
\end{figure}

It is possible to predict $D_\lambda$ given $\mathbf{A}_c$ and $\lambda_1$ using Eqs.~\eqref{eq:lyapunov-spectrum} and~\eqref{eq:fractal-dimension}. Figure~\ref{fig:fractal-dimension} shows the comparison of the numerically computed $D_\lambda$ (red, circles) with the theoretical values (blue, squares) over a broad range of $\xi_d$. The agreement between the numerical results and the theoretical prediction is excellent over the entire range shown.

Figure~\ref{fig:fractal-dimension} shows that the fractal dimension decreases with increasing $\xi_d$. This indicates that the amount of disorder in the dynamics decreases, on average, as the length scale of the spatial coupling increases.  This is despite the fact that $\lambda_1$ is not sensitive to $\xi_d$ over the range shown in Fig.~\ref{fig:lyap-with-xid}. The variation of $D_\lambda$ with $\xi_d$ can be traced to the variation of the entire Lyapunov spectrum with $\xi_d$ as illustrated in Figs.~\ref{fig:lyapunov-spectrum}-\ref{fig:lyapunov-spectrum-comparison}.  The solid line in Fig.~\ref{fig:fractal-dimension} is a curve-fit through the numerical results of the form $D_\lambda \!=\! 170.17 \xi_d^{-1/2} \!-\! 33.08$. This yields that $D_\lambda$ depends upon the length scale of the coupling as $\xi_d^{-1/2}$. For $\xi_d \!\ge\! 8$ the dynamics are periodic in time and the fractal dimension does not exist. 

We highlight that the dimension of the dynamics is large for small values of $\xi_d$, for example $D_\lambda \!\approx\! 138$ for $\xi_d \!=\! 2$. This indicates that a numerical calculation of $D_\lambda$ would require the simultaneous evolution of the nonlinear dynamics and $\sim 140$ instances of the tangent space equations.  However, using the theoretical approach $D_\lambda$ can be estimated with knowledge of $\lambda_1$ and $\Lambda_k$. Given an analytical expression for $\Lambda_k$, for any form of linear spatial coupling, a fundamental understanding of how $D_\lambda$ will vary with the spatial coupling can be obtained. We note that if $\lambda_0$ is used as the Lyapunov exponent for $k\!=\!1$ in Eq.~\eqref{eq:lyapunov-spectrum}, this provides an even more accessible approximation which gets the shape of the curve correct while over predicting the values of the fractal dimension since $\lambda_0 > \lambda_1$ for the spatial coupling we consider.  

\subsection{Covariant Lyapunov Vectors}
\label{section:clvs}

\begin{figure}[h!]
\begin{center}
\includegraphics[height=0.5\textwidth]{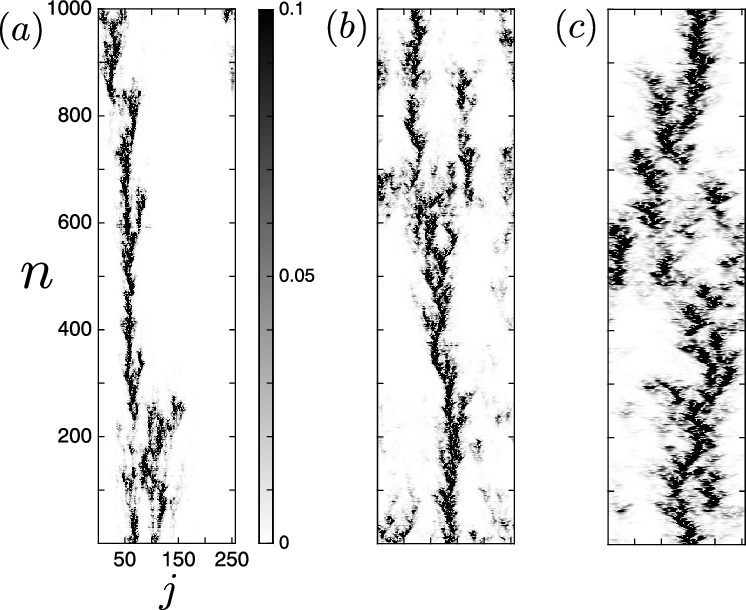}
\end{center}
\caption{Spacetime plots of the magnitude of the components of the leading CLV, $\vec{v}_1^{(n)}$, for ($a$) $\xi_d \!=\! 1$, ($b$) $\xi_d \!=\! 2$, ($c$) $\xi_d \!=\! 6$. The discrete time is $n$ and $j$ is the component of the $N$-dimensional CLV. All panels use the same color bar.}
\label{fig:clvs}
\end{figure}

We now explore the CLVs and discuss their fundamental connection with the eigenvectors of the coupling matrix. Figure~\ref{fig:clvs} shows spacetime plots of the magnitude of the components of the leading CLV, $\vec{v}_1^{(n)}$, for three different values of $\xi_d$. All three plots use the same gray scale where dark is a large value and white is a small value. The abscissa, $j$, indicates the component of the leading CLV, and the ordinate is the discrete time $n$. 

The spacetime plots indicate that $\vec{v}_1^{(n)}$ is spatially localized. The localization of Lyapunov vectors has been reported elsewhere for CMLs~\cite{pikovsky:1998,takeuchi:2011} and for extended fluid systems such as Rayleigh-B\'enard convection~\cite{xu:2016,xu:2018}. Figure~\ref{fig:clvs} indicates that the degree of spatial localization of the leading CLV decreases as the length scale of the spatial coupling increases.

\begin{figure}[h!]
\begin{center}
\includegraphics[width=3.8in]{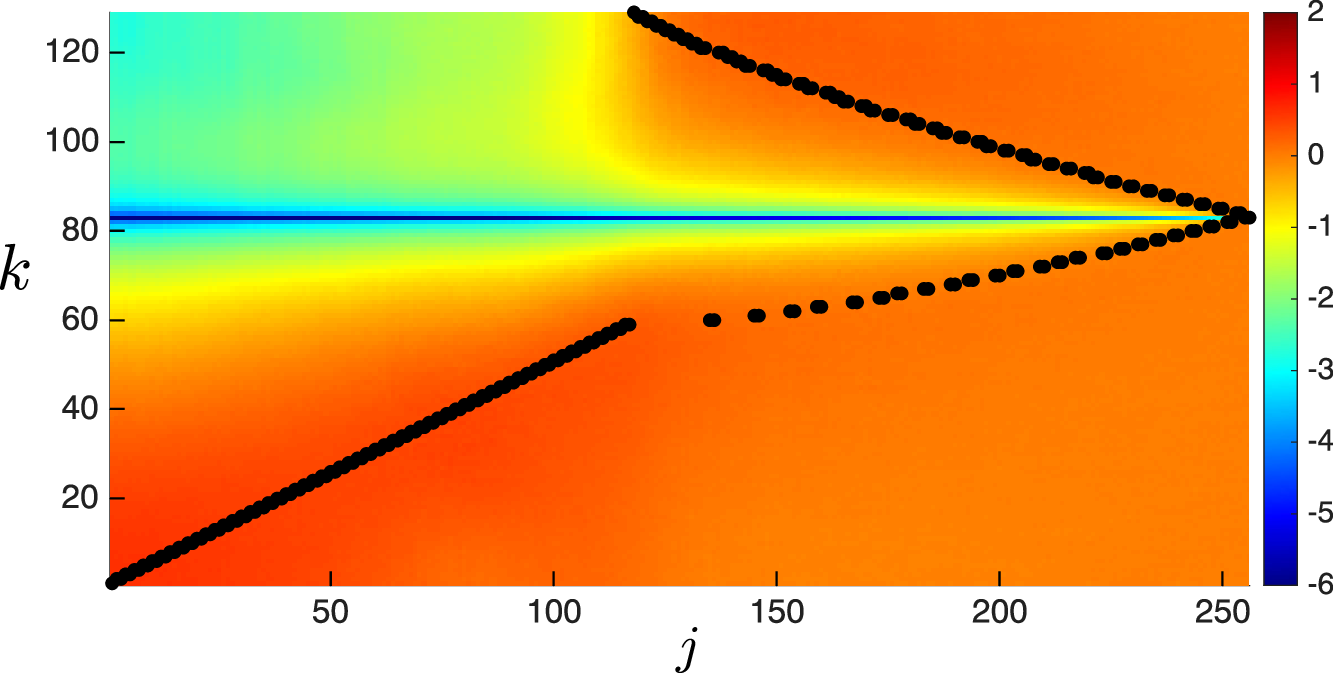}
\end{center}
\caption{The spatial power spectrum of the CLVs for $\xi_d \!=\! 1$. Color contours are of the time average of the magnitude of the spatial power spectrum of the CLVs, $\langle |\hat{v}_j|^2 \rangle$, using the $\log_{10}$ scale with red (large) and blue (small) where the angle brackets indicate a time average. The Lyapunov index is $j$ where $j \!=\! 1,2, \ldots, N$ and $k$ is the integer wavenumber. The black circles are the wavenumbers of the sorted eigenvectors $\vec{w}_j$ of the coupling matrix when sorted in descending order based upon the magnitude of their corresponding eigenvalue $|\Lambda_j|$. The Lyapunov exponent closest to zero is $\lambda_{80} \!\approx\! 6.37 \!\times\! 10^{-4}$.} 
\label{fig:sps1}
\end{figure}

The length scales that contribute to the CLVs can be captured using a spatial power spectrum. Figure~\ref{fig:sps1} shows the time-averaged spatial power spectrum of the CLVs $\langle|\hat{v}_j |^2 \rangle$ for $\xi_d \!=\! 1$ where the hat notation indicates a discrete Fourier transform and the angle brackets indicate an average over time. The Lyapunov index of the sorted CLVs is $j$, where $j \!=\!1$ is the leading CLV which is then followed by the remainder of the CLVs for $j=2,\ldots,256$. The integer wavenumber of the CLV is $k$. A vertical slice of Fig.~\ref{fig:sps1} is the time average of the spatial power spectrum of an individual CLV. In all of the spatial power spectra, the time averaging is over $5 \!\times\! 10^4$ time steps.  The magnitude of the power spectrum is indicated by the color contours using the $\log_{10}$ scale on the right. 

The lower left region of Fig.~\ref{fig:sps1}, where $j \!\le\! 117$ and $k \!\le\! 59$, has significant contributions as indicated by the overall red color. For larger wavenumbers, the CLVs have very little contribution as indicated by the dark blue region.  This indicates that the CLVs with the largest positive Lyapunov exponents have significant contributions at smaller wavenumbers corresponding with larger spatial structures.

The unsorted eigenvectors $\vec{w}_{k'}$ of $\mathbf{A}_c$ are unit-vector Fourier modes with integer wavenumber. The eigenvectors are 
\begin{equation}
\vec{w}_{k'} = \frac{1}{\sqrt{N}} \cos \left[ \frac{\pi (k'-1) m}{N} \right] 
\label{eq:evec1}
\end{equation}
for $k' \!=\! 1, \ldots, N/2 \!+\! 1$, and  
\begin{equation}
\vec{w}_{k'} = \frac{1}{\sqrt{N}} \sin \left[ \frac{\pi (k' - \frac{N}{2}) m }{N} \right]
\label{eq:evec2}
\end{equation}
for $k' \!=\! N/2 \!+\! 2, \ldots, N$.  The $\vec{w}_{k'}$ are $N$ dimensional vectors where $m \!=\!1,2, \ldots N$. The eigenvectors, given by Eqs.~(\ref{eq:evec1})-(\ref{eq:evec2}), each have the unsorted eigenvalue $\Lambda_{k'}$ given by Eq.~\eqref{eq:eigenvalues-unsorted}.

The leading eigenvector, for $k' \!=\! 1$, is a constant. For increasing $k'$, this is followed by the cosine modes and then the sine modes. The integer wavenumber of each eigenvector is given by the term in parentheses in the numerator of the cosine or sine term. The leading eigenvector is a constant vector with values $N^{-1/2}$, there are $N/2 \!-\!2$ eigenvector pairs of sines and cosines with the same wavenumber, and a single eigenvector with a wavenumber of $N/2\!+\!1$.

In Fig.~\ref{fig:sps1}, the spatial power spectrum of the eigenvectors are indicated by the black circle symbols. For the eigenvectors shown in Fig.~\ref{fig:sps1}, $j$ is the index of the sorted eigenvector and $k$ is the integer wavenumber of the eigenvector. Since the eigenvectors are Fourier modes, their power spectrum yields a contribution  only at the value of their integer wavenumber (indicated by the black circles). A close inspection yields that the eigenvectors appear in pairs, except for $j \!=\! 1$ and $j \!=\! 129$, as expected.

There is a significant connection between the wavenumber of the eigenvector and the wavenumbers of the CLVs where large contributions are shown in red. For $j \!\le\! 117$ the wavenumbers of the eigenvectors increases linearly in pairs. For $j \!>\! 117$ the wavenumbers of the eigenvectors are no longer increasing linearly and there is significant mixing of wavenumbers with increasing $j$.

This can be traced to the reordering of the eigenvectors by the magnitude of their corresponding eigenvalue from largest to smallest as required by the calculation of the Lyapunov exponent in Eq.~\eqref{eq:lyapunov-spectrum}.  The blue horizontal stripe, centered at $k \!=\! 83$, represents the wavenumber of the eigenvector with $|\Lambda_j|$ closest to zero.  For reference, the Lyapunov exponent whose value is closest to zero is $\lambda_{80} \!\approx\! 6.37 \!\times\! 10^{-4}$. This illustrates that the region $j \le 117$ includes all of the CLVs with positive Lyapunov exponents as well as some of the CLVs with negative Lyapunov exponents (see Fig.~\ref{fig:lyapunov-spectrum}(a)). 

\begin{figure}[h!]
\begin{center}
\includegraphics[width=3.5in]{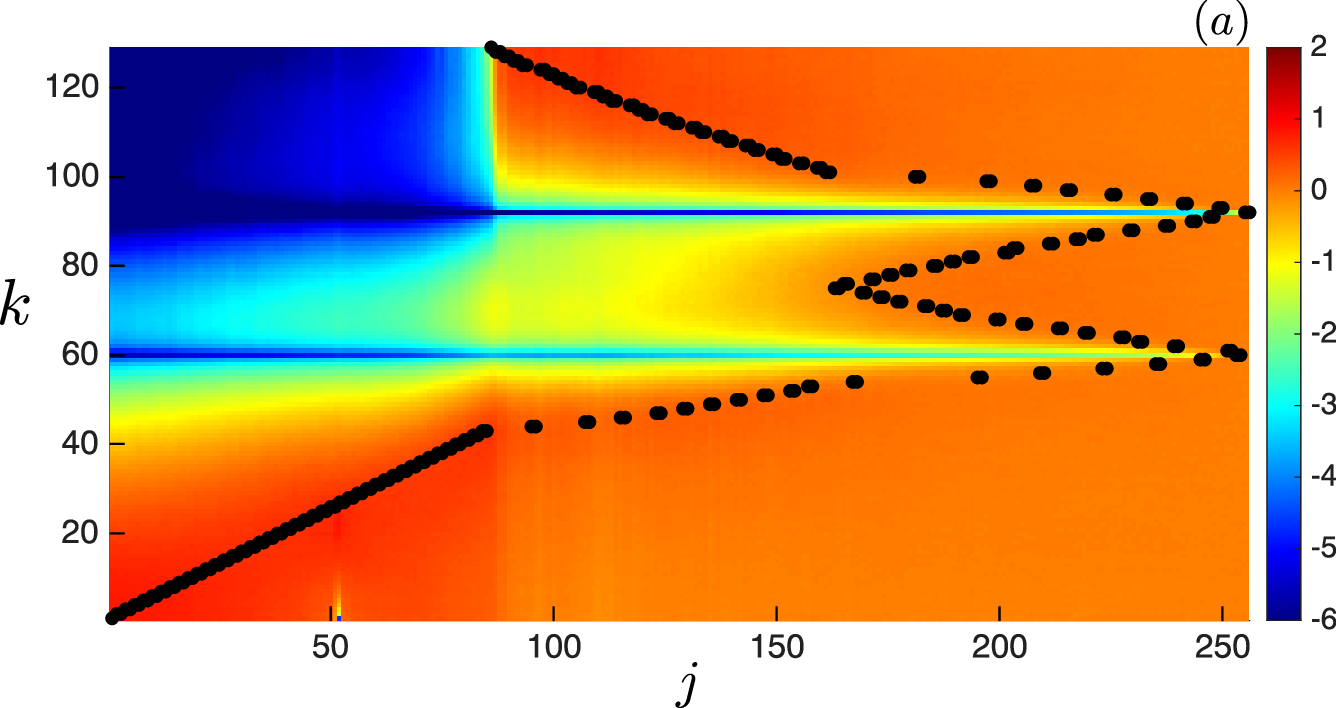}  
\hspace{0.1cm}
\includegraphics[width=3.32in]{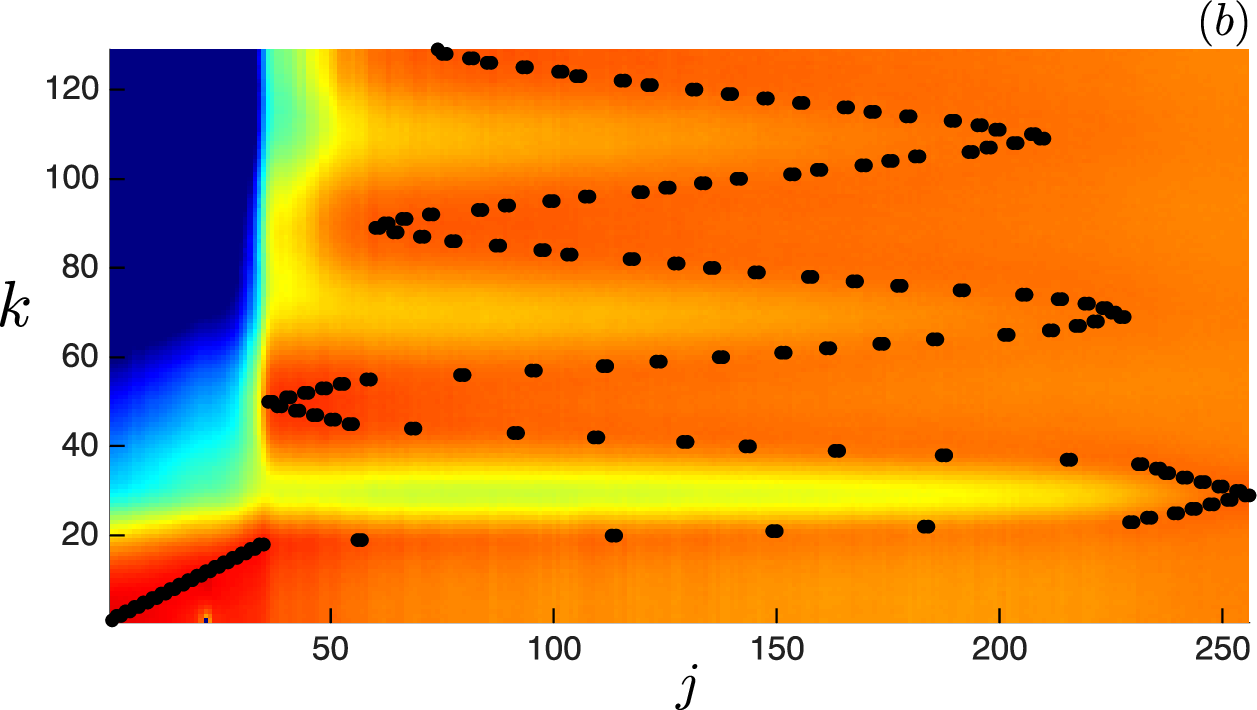} 
\end{center}
\caption{The spatial power spectrum of the CLVs using the same conventions as Fig.~\ref{fig:sps1}. ($a$) $\xi_d \!=\! 2$, the Lyapunov exponent closest to zero is $\lambda_{52} \!\approx\! -4.08 \times 10^{-6}$. ($b$) $\xi_d \!=\! 6$, the Lyapunov exponent closest to zero is $\lambda_{22} \!\approx\! -1.31 \times 10^{-5}$.  Both panels use the color bar shown.} 
\label{fig:sps2}
\end{figure}

The connection between the CLVs and the eigenvectors in the spatial power spectrum is present for all values of $\xi_d$. Figure~\ref{fig:sps2} shows the spatial power spectra for ($a$) $\xi_d \!=\!2$ and ($b$) $\xi_d\!=\!6$.

The spatial features of the CLVs reflect the wavenumbers of the Fourier-mode eigenvectors when they are sorted by $|\Lambda_k|$ in descending order. As a result, this yields an interesting variation of the eigenvector wavenumber with increasing Lyapunov index.  The eigenvectors with the smallest wavenumber, corresponding to the largest spatial structures, occur in pairs that increase linearly with the Lyapunov index. This is followed by a mixing of the wavenumbers for increasing Lyapunov index.  The wavenumber mixing occurs when $\Lambda_k$ becomes negative, which yields the cusp-like features in the eigenvalue spectra (for example, see Fig.~7 in Ref.~\cite{raj:2024}). The horizontal stripes of very small value (blue) occur at the wavenumbers of the eigenvectors whose eigenvalues are close to zero.

\subsection{Violation of the Domination of Oseledets Splitting}
\label{section:vdos}

The spectrum of Lyapunov exponents $\lambda_i$ are guaranteed to be in descending order by Oseledets Multiplicative Ergodic Theorem~\cite{oseledec:1968}. However, over finite intervals of time, the finite-time CLEs $\tilde{\lambda}_i$ may not follow this strict ordering. This is referred to as violations of the domination of Oseledets splitting (DOS). These violations indicate instances of time when the CLVs are nearly transversal~\cite{pugh:2003,bochi:2005}. When CLVs are nearly transversal, a small perturbation to the dynamics in their direction could affect the growth of both vectors. When this occurs, the CLVs are often described as entangled. It is important to note that the violation of DOS calculation must be conducted with the CLEs and not with the finite time Lyapunov exponents based upon the OLVs~\cite{takeuchi:2011}.

The fraction of time when such a violation occurs can be quantified by $\nu_{k_1,k_2}^\tau$ where $k_1$ and $k_2$ are the Lyapunov indices of two CLVs and $\tau$ is the finite interval of time used to calculate $\tilde{\lambda}_{k_1}$ and $\tilde{\lambda}_{k_2}$. The difference between two finite-time CLEs, over the time interval $\tau$, is $\Delta \tilde{\lambda}_{k_1,k_2}^\tau \!=\! \tilde{\lambda}_{k_1} \!-\! \tilde{\lambda}_{k_2}$. A violation of the Oseledets splitting occurs when $\Delta \tilde{\lambda}_{k_1,k_2}^{\tau} \!<\! 0$ for $k_1 \!>\! k_2$. For very long intervals of time, $t \!\gg\! \tau$, there will be many values of $\Delta \tilde{\lambda}_{k_1,k_2}^{\tau}$ computed.  The fraction of the computed values of $\Delta \tilde{\lambda}_{k_1,k_2}^{\tau}$ for which a violation occurred can be expressed as $\nu_{k_1,k_2}^\tau = \langle 1 - \mathcal{H} (\Delta \tilde{\lambda}_{k_1,k_2}^{\tau})\rangle_t$ where $\mathcal{H}$ is the Heaviside operator and $\langle \cdot \rangle_t$ indicates a time average~\cite{takeuchi:2011}.
\begin{figure}[h!]
\begin{center}
\includegraphics[width=2.75in]{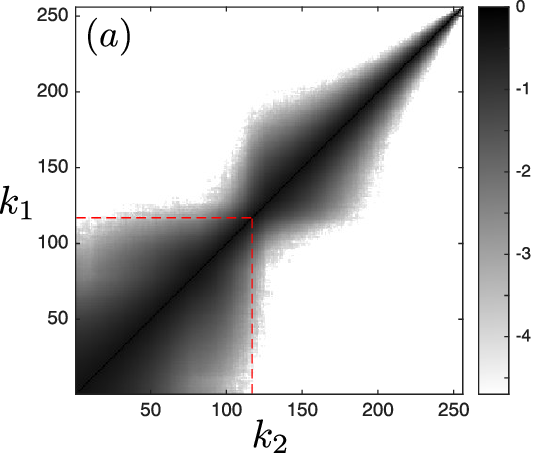} \hspace{0.4cm}
\includegraphics[width=2.75in]{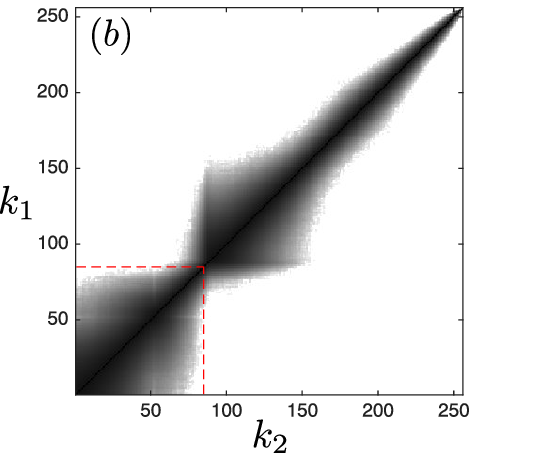}  \\ \vspace{0.4cm}
\includegraphics[width=2.75in]{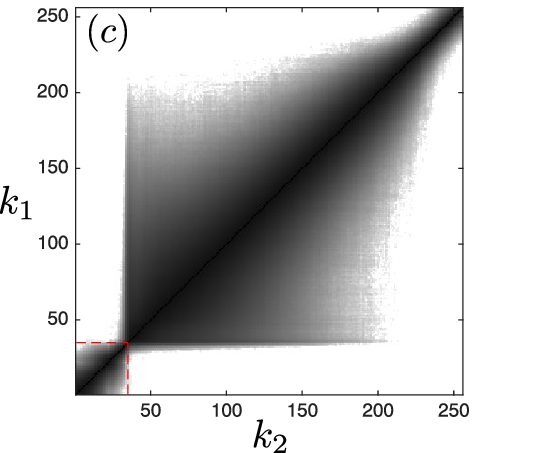}
\end{center}
\caption{Violation of DOS, $\nu_{k_1,k_2}^\tau$, for ($a$) $\xi_d\!=\!1$, $k^*\!=\!117$; ($b$) $\xi_d\!=\!2$, $k^*\!=\! 85$; ($c$) $\xi_d \!=\! 6$, $k^*\!=\!35$. $k_1$ and $k_2$ are Lyapunov indices, all panels use the $\log_{10}$ scale indicated by the color bar. In all of our calculations, we have used $\tau\!=\!5$. Red dashed lines indicate the index $k^*$ corresponding to the index of the last eigenvector in the linear regime shown by the black symbols in Fig.~\ref{fig:sps1}-\ref{fig:sps2} which exists for $1 \!\le\! j \!\le\! k^*$.} 
\label{fig:vdos}
\end{figure}

Figure~\ref{fig:vdos} shows the variation of $\nu_{k_1,k_2}^\tau$ for three values of $\xi_d$ using the $\log_{10}$ gray scale indicated by the color bar. Dark regions represent pairs of CLVs, given by indices $k_1$ and $k_2$, that experience violations for a significant amount of time. White regions represent pairs without violations.  The diagonal from the bottom left to the top right represents pure violation since $k_1\!=\!k_2$ and therefore $\tilde{\lambda}_{k_1}$ is never less than $\tilde{\lambda}_{k_2}$. The figures are symmetric about this diagonal by construction. In our calculations, we have used $\tau \!=\! 5$. We have confirmed that the general features illustrated in Fig.~\ref{fig:vdos} do not change in a significant way when using values of $\tau$ over the range $2 \!\le\! \tau \!\le\! 15$.

Figure~\ref{fig:vdos}($a$) illustrates $\nu_{k_1,k_2}^\tau$ for $\xi_d \!=\! 1$. Overall, there is a significant amount of violation and all of the CLVs are tangled with their neighbors. There is an interesting structure in the violations of DOS plot. The most violations occur at the smaller Lyapunov indices and the smallest amount of violation occurs for the largest values of $k_1$ and $k_2$.

Further insight into the structure of the violation of DOS plot can be gained by again appealing to the eigenvectors of $\mathbf{A}_c$. The red dashed lines shown in Fig.~\ref{fig:vdos} indicate the index range where the eigenvector pairs exhibit a linear dependence upon wavenumber as shown in Fig.~\ref{fig:sps1}. The largest wavenumber in this group is denoted as $k^*$. The region enclosed by the red dashed lines contains all of the CLVs with positive Lyapunov exponents and a small portion of the CLVs with small negative Lyapunov exponents. The value of $k^*$ is a good prediction for the fractal dimension $D_\lambda$ as shown by the green triangles in Fig~\ref{fig:fractal-dimension}.  In addition, this region contains the eigenvectors in the spectrum with the smallest wavenumbers which occur in pairs. 

The region contained within the red dashed lines decreases with increasing values of $\xi_d$. It is interesting to draw attention to the fact that chaos essentially disappears for $\xi_d \!\ge\! 8$ as shown in Fig.~\ref{fig:xic}($b$). This corresponds with the vanishing of the red shaded region in Fig.~\ref{fig:vdos}. This suggests that the presence of unmixed eigenvectors with small wavenumbers and their connection with the CLVs with positive Lyapunov exponents may be important for the chaotic dynamics.

The violation of DOS shows a signature of the eigenvectors.  The violations at small wavenumbers are fully contained in the region outlined by the red dashed lines. Despite being in a regime where the eigenvectors are pairs of Fourier modes of linearly increasing wavenumber, the violations contained in the region bounded by the red dashed lines extend to include all of the corresponding CLVs as indicated by the gray contours which nearly all occur in this bounded region. The CLVs with larger indices that extend outside of this range do not contribute to the violations. 

\section{Conclusion}
\label{section:conclusion}

We have explored the spatiotemporal chaos of a large 1D lattice of coupled quadratic maps as the extent of the spatial coupling is increased. Using the CLVs, we have computed the Lyapunov spectrum, the fractal dimension, the spatial power spectrum of the CLVs, and the violation of DOS as $\xi_d$ is varied. We find that the leading Lyapunov exponent $\lambda_1$ is insensitive to $\xi_d$ for chaotic dynamics until the dynamics become periodic for large $\xi_d$. For the range of $\xi_d$ yielding chaotic dynamics, the fractal dimension decreases as $\xi_d^{-1/2}$ for increasing values of $\xi_d$. The spatial power spectrum of the CLVs indicate significant contributions at small wavenumbers, suggesting the importance of larger length scale structures. Quantifying the violation of DOS yields that all of the CLVs are entangled for all of the conditions that we explore.

Important insights into the chaotic dynamics can be gained using the eigenvectors and eigenvalues of the spatial coupling matrix $\mathbf{A}_c$. Using only the eigenvalues and a computation of $\lambda_1$, it is possible to analytically predict the entire Lyapunov spectrum and therefore the fractal dimension.  Furthermore, the eigenvectors yield insight into the general spatial structure of the CLVs and their entanglement which indicates the presence of two different regimes.  The leading eigenvectors, when sorted, have small wavenumbers whose values align with the wavenumbers where the leading CLVs exhibit significant magnitude in the spatial power spectrum. The wavenumber of the eigenvectors in this range increases linearly with their index. The CLVs in this range are highly entangled with one another.  This is then followed by eigenvectors of mixed wavenumbers for increasing index which is clearly reflected in the spatial power spectrum of the CLVs. The CLVs in this range are much less entangled. The index dividing these two regimes is approximately equal to the fractal dimension of the dynamics. In light of these results, it is possible to predict very general and fundamental features of the CLVs using only the eigenvectors of $\mathbf{A}_c$.

It is important to highlight that the analytical insights into the chaotic dynamics, and into the CLVs of the CMLs that we present here, are independent of the mapping function that is used and depend only upon the spatial coupling that is present. It would be very interesting to explore this quantitatively using different nonlinear mapping functions although we have not pursued that here.  It would be insightful to explore the findings presented here in detail for more complex dynamical systems, such as a nonlinear partial differential equation. We anticipate that these insights into the basic structure of the CLVs and CLEs will be of interest to researchers working on nonlinear and chaotic dynamical systems in a broad range of fields.

~\vspace{0.2cm}

\noindent Acknowledgments: We acknowledge support from NSF fund number CMMI-2138055. Portions of the computations were performed using the Advanced Research Computing (ARC) center at Virginia Tech.

\bibliographystyle{ws-ijbc}

\end{document}